# Stochastic resonance in a quasi-two-dimensional superlattice. II.

G.M. Shmelev[1], E.M. Epshtein[2], A.S. Matveev[1]

[1]Volgograd State Pedagogical University, 400131, Volgograd, Russia

[2]Institute of Radio Engineering and Electronics of the Russian Academy of Sciences, 141190, Fryazino, Russia

**Abstract**. A fluctuation theory is presented for the nonequilibrium second order phase transition in a quasi-two-dimensional electron gas. A transverse (with respect to the current through the sample) spontaneous electric field as an order parameter and a driving longitudinal field as a control parameter are used. In addition to the earlier results, the intrawell dynamics is taken into account. Non-monotonous behavior of the periodic signal gain as a function of the noise power (stochastic resonance) is predicted.

## 1. Introduction

Low-dimensional semiconductor structures (quantum wells, wires, rings, dots, superlattices, etc.) draw great attention now. Such systems manifest nonlinear and nonequilibrium properties in rather low fields. Theory of nonequilibrium phase transitions (NPT) is another fast-developing branch of modern physics. Therefore, merging these two branches and appearing works devoted to NRT in low-dimensional structures look quite naturally.

Recently, a principal possibility has been established for existence of nonequilibrium phase transitions (NPT) in a quasi-two-dimensional electron gas (2DEG) under square lateral modulation [1 – 5]. Such 2DEG exists in a quasi-two-dimensional semiconductor superlattice (2SL) with a crystal potential periodically modulated in two dimensions whereas the electron motion in the third direction is confined. A transverse (relative to the current $j_x$ flowing in the sample) spontaneous electric field $E_y$ (or a longitudinal magnetic field $H_x$) appears as an order parameter. As controlling parameters, the applied field $E_x$, the sample temperature $T$ or the external circuit finite resistance in the $y$ direction can be taken. Under such conditions, the nonequilibrium electron gas displays properties typical of ferroelectrics (or ferromagnets). The spontaneous symmetry breakdown and appearance of the field $E_y$ give a simple example of self-organization in nonequilibrium electron gas; moreover, the model in consideration is exactly soluble one.

As in the equilibrium phase transitions, the role of fluctuations rises substantially near the NPT. Particularly, the synergetic potential bistability (see below) indicates the possibility of a stochastic resonance due to the noise-induced transitions between the potential minima.

The stochastic resonance phenomenon was discovered in 1981 as a possible explanation of the periodical changes of the global climate [6]. Since then the stochastic resonance was predicted and found in very different branches, namely, in laser, chemical and biological systems, in economical models [7, 8], etc. (Some authors consider the term "stochastic resonance" as not good enough [9]). Under simultaneous acting harmonic signal and noise on a bistable system, the signal amplitude can enhance and the gain as a function of the noise power has a maximum under some optimal value of the power. Such a behavior of the gain and the other integral characteristics is explained in terms of the Kramers rate $r_k$, the mean frequency of the system transitions from one stable state to other one. The maximum mentioned takes place when the signal frequency $\Omega$ is of the same order of magnitude as $2r_k$.

In Refs. [10, 11] a response of the nonequilibrium 2DEG to a weak electric periodical signal along $y$ axis was calculated as a function of the driving field $E_x$ and temperature. In present work, effect of the intrawell dynamics [7, 8] on the stochastic resonance in 2SL is investigated.

## 2. Distribution function of the spontaneous transverse electric field in quasi-two-dimensional superlattice

We consider a 2SL with electron energy spectrum

$$\varepsilon(\vec{p}) = \varepsilon_0 - \Delta \cos\left(\frac{p_x d}{\hbar}\right) \cos\left(\frac{p_y d}{\hbar}\right), \qquad (1)$$

where **p** is quasimomentum, $2\Delta$ is the conduction miniband width. The spectrum (1) has a standard form of the tight-binding approximation, $x$ and $y$ axes are directed at an angle of 45° relative to the principal axes of a simple square lattice with period $a$, $d = a/\sqrt{2}$. The electron motion normal to the 2SL surface (along $z$ axis) is "frozen out" because of quantum confinement (the 2SL thickness is $\sim 10^{-6} - 10^{-7}$ cm). A quasi-classical situation is considered when $2\Delta \gg \hbar/\tau, |eE|d$ ($\tau$ being electron mean free path) and the distribution function can be found by solving the Boltzmann kinetic equation with a collision integral within the $\tau$-approximation. We use a constant relaxation time. Such an approach reflects the results and analysis of current experiments with SLs [12]. Besides, no principally new results concerned NPT appear beyond that approximation [3, 4].

Let us introduce dimensionless variables by the following redefinition: $\mathbf{E}/E_0 \to \mathbf{E}$, $\mathbf{j}/j_0 \to \mathbf{j}$, $t/\tau_0 \to t$, $\Omega\tau_0 \to \Omega$, where $E_0 = \hbar/ed\tau$, $j_0 = \sigma_0 E_0$, $\tau_0 = \varepsilon/(4\pi\sigma_0)$, $\sigma_0 = e^2 n\Delta d^2 \tau/\hbar^2$, $n$ is electron density, $\varepsilon$ is dielectric constant.

Using the distribution function, we obtain the current density [1]:

$$j_y = C_{11} E_y (1 + E_y^2 - E_x^2) W^{-1}, \tag{2}$$

$$W(E_x^2, E_y^2) = (1 + E_x^2 + E_y^2)^2 - 4E_x^2 E_y^2, \tag{3}$$

where

$$C_{11}(T) = \left\langle \cos\frac{p_x d}{\hbar} \cos\frac{p_y d}{\hbar} \right\rangle_0$$

is an average over the equilibrium distribution. A similar expression for $j_x$ is obtained from Eq. (2) by interchange $x \leftrightarrow y$.

Let the sample be opened in $y$-direction, i.e.,

$$j_y = 0. \tag{4}$$

Then it follows from Eqs. (2) – (4) that

$$E_y \equiv E_{ys} = \begin{cases} 0, & 0 \leq |E_x| < 1 \\ \pm\sqrt{E_x^2 - 1}, & |E_x| \geq 1 \end{cases} \tag{5}$$

The zero solution is unstable against small fluctuations of the electric field at $|E_x| \geq 1$. The solutions (5) satisfy the stability condition [13]

$$\left.\frac{\partial j_y}{\partial E_y}\right|_{E_y = E_{ys}} > 0. \tag{6}$$

In terms of synergetic potential [1, 2],

$$\Phi(E_x^2, E_y^2, T) = \int j_y dE_y + \text{const}. \tag{7}$$

Equations (4) and (6) take the form

$$\left.\frac{\partial \Phi}{\partial E_y}\right|_{E_y=E_{ys}} = 0, \quad \left.\frac{\partial^2 \Phi}{\partial E_y^2}\right|_{E_y=E_{ys}} > 0. \tag{8}$$

In our case,

$$\Phi = \tfrac{1}{4} C_{11} \ln W + \text{const}. \tag{9}$$

The conditions (8) determine a minimum of the potential $\Phi$ (the latter becomes two-well one at $|E_x| > 1$, see Fig. 1). Thus, Eqs. (5) and (8) mean that a NPT of the second order takes place at $|E_x| = 1$ that has been considered in Refs. [1 – 5].

In an unsteady case, the open-circuit condition in $y$ direction is to be written, instead of (4), as

$$j_y + \frac{dE_y}{dt} = 0. \tag{10}$$

In view of Eq. (7), Eq. (10) takes form

$$\frac{dE_y}{dt} = -\frac{\partial \Phi}{\partial E_y}. \tag{11}$$

Note that Eq. (11) is nothing but a phenomenological Landau – Khalatnikov equation [14] that is obtained here by a natural way without any assumption of closeness to equilibrium. Adding a random (fluctuation) current density $\delta j_y$ to the right-hand side of Eq. (11), we obtain the Langevin equation

$$\frac{dE_y}{dt} = -\frac{\partial \Phi}{\partial E_y} - \delta j_y(t). \tag{12}$$

To consider the thermal (additive and non-correlated) noise we take the known expression for the current correlation function [14]

$$\langle \delta j_y(t) \delta j_y(t') \rangle = 2\theta \delta(t - t'), \tag{13}$$

where $\theta = 4\pi C_{11} kT / \varepsilon V E_0^2$, $V$ is the system volume, $k$ is the Boltzmann constant.

Using standard formalism [15], we write a Fokker – Planck equation equivalent to Eq. (12) for the distribution function $f(E_y, t)$ of the random quantity $E_y$:

$$\frac{\partial f}{\partial t} = \frac{\partial}{\partial E_y}(j_y f) + \theta \frac{\partial^2 f}{\partial E_y^2}. \tag{14}$$

It follows from Eq. (14) that the steady state distribution function is

$$\bar{f}(E_x^2, E_y^2, T) = A e^{-\frac{\Phi}{\theta}} \quad (15)$$

under the condition $\bar{f}\big|_{E_y^2=\infty} = 0$.

The constant $A = A(E_x^2, T)$ is determined by a normalization condition

$$\int_{-\infty}^{+\infty} \bar{f}(E_x^2, E_y^2, T) dE_y = 1. \quad (16)$$

Taking into account Eq. (9), the function (15) can be written in the form of

$$\bar{f}(E_x^2, E_y^2, T) = A [W(E_x^2, E_y^2)]^{-\alpha} \quad (17)$$

where $\alpha = \varepsilon V E_0^2 / (16 \pi kT)$. The extrema of the steady distribution function at $E_x$ = const are the values $E_y = E_{ys}$ determined by Eq. (5).

The mean-square value of the transverse field fluctuation

$$\langle E_y^2 \rangle = \int_{-\infty}^{+\infty} E_y^2 \bar{f}(E_x^2, E_y^2, T) dE_y \bigg/ \int_{-\infty}^{+\infty} \bar{f}(E_x^2, E_y^2, T) dE_y, \quad (18)$$

in our case, is

$$\langle E_y^2 \rangle = \frac{(1 + E_x^2)}{4\alpha - 3} \cdot \frac{P_{1-\alpha}^{\frac{1-\alpha}{2}}(x)}{P_{-\alpha}^{\frac{1-\alpha}{2}}(x)}, \quad (19)$$

where $P_\nu^\mu(x)$ is the associated Legendre function of the first kind and $x = (1 - E_x^2)/(1 + E_x^2)$.

In absence of the driving field ($E_x = 0$) it follows from Eq. (19) that

$$\langle E_y^2 \rangle = 1/(4\alpha - 3). \quad (20)$$

This result differs somewhat from the Nyquist formula [16] ($\langle E_y^2 \rangle = 1/4\alpha$) which means that the theory proposed is valid at $\alpha \gg 3/4$.

Let us make some numerical estimates. At $d = 10^{-6}$ cm and $\tau = 10^{-12}$ s we have $E_0 \approx 600$ V/cm. At $\varepsilon = 10$ and $T = 100$ K the condition $\alpha \gg 1$ is equivalent to the condition $V \gg 10^{-14}$ cm$^{-3}$. If the sample length in $x$ direction is $l_x = 1$ cm and the thickness is $l_z = 10^{-6}$ cm then we have a condition $l_y \gg 10^{-8}$ cm that fulfils well for the electric field domains ($l \sim 10^{-4} - 10^{-5}$ cm [13]).

The average value $\langle E_y^4 \rangle$ is calculated by similar way. We obtain

$$\left\langle E_y^4 \right\rangle = \frac{3\left(1+E_x^2\right)^2}{16\alpha^2 - 32\alpha + 15} \cdot \frac{P_{2-\alpha}^{\frac{1}{2}-\alpha}(x)}{P_{-\alpha}^{\frac{1}{2}-\alpha}(x)}. \tag{21}$$

At $E_x^2 = 1$ we have $\left\langle E_y^4 \right\rangle = \alpha^{-1}$; the coefficient of excess $\gamma = \left\langle E_y^4 \right\rangle / \left(\left\langle E_y^2 \right\rangle\right)^2 - 3$ [17] in that point is $\gamma = -0.812$. The |γ| closeness to 1 shows that the fluctuations in the critical range are not only large but also non-Gaussian ones.

### 3. Stochastic resonance

Stochastic resonance [7, 8] (or stochastic filtration [9]) determines a class of phenomena in which the response of a nonlinear system to a weak external signal rises substantially with increasing in the noise power in the system. The integral characteristics of the process (e.g., gain or signal-to-noise ratio) have a pronounced maximum at some optimal noise level. Such behavior is interpreted by means of average time $\tau_k$ of noise-induced transition of the two-well system from one stable state to another one (the states with $E_{ys} = \sqrt{E_x^2 - 1}$ and $E_{ys} = -\sqrt{E_x^2 - 1}$, in our case).

The rate $r_k$ can be expressed via the stationary distribution function [15]:

$$r_k^{-1} = \frac{1}{\theta} \int\limits_{-\sqrt{E_x^2 - E_0^2}}^{+\sqrt{E_x^2 - E_0^2}} \frac{dE_y}{\bar{f}(E_x^2, E_y^2, T)} \int\limits_{-\infty}^{E_y} \bar{f}(E_x^2, E_y'^2, T) dE_y'. \tag{22}$$

In calculations with Eq. (22), so called parabolic approximation is used frequently [7, 8, 15]. This approximation is not necessary here, since we have an exact (in scope of the model considered) distribution function (15), (17).

Let a harmonic signal add to the situation considered, e.g., by means of a plane capacitor with sinusoidal voltage on its plates perpendicular to *y* axis. Correspondingly, a term $C_{11} F \cos(\Omega t)$ is added to the right-hand side of Eq. (12) where *F* is the amplitude of ac field applied to the sample.

The power gain factor is defined as squared modulus of the system linear susceptibility ($\eta(\omega) = |\chi(\omega)|^2$). According to [18, 19], we have the susceptibility at $\alpha \gg 1$ with intrawell dynamics taking into account:

$$\chi(\omega) = \frac{1}{\theta} 2 r_k \left\langle E_y^2 \right\rangle (2 r_k - i\omega)^{-1} + \left(\Phi''(E_{ys}) - i\omega\right)^{-1}; \tag{23}$$

the second summand corresponds to intrawell dynamics.

Gain factor is shown in Figs. 1 and 2 for 2SL with the following parameters: $\Delta = 2 \times 10^{-2}$ eV, $d = 1 \times 10^{-6}$ cm, $\tau = 1 \times 10^{-12}$ s, $n = 1 \times 10^{16}$ cm$^{-3}$, $V = 2 \times 10^{-13}$ cm$^3$, $\varepsilon = 10$; the field $E_x$ is expressed in units of $E_0 \approx 600$ V/cm and the frequency $\Omega$ in units of $\tau_0^{-1} = 4\pi\sigma_0/\varepsilon$ ($\tau_0^{-1} \approx 9.3 \times 10^{13}$ s$^{-1}$). The dotted line does not consider intrawell dynamics, the solid lines show the total gain factor with untrawell and global dynamics, the dashed line presents the function $\left(\Phi_n''^2 + \omega^2\right)^{-1}$. As was to be expected, including the intrawell dynamics gives nonzero gain factor $\eta \approx \left(\Phi_n''^2 + \omega^2\right)^{-1}$ [8] in the limit of $\theta \to 0$ ($\alpha \to \infty$). A similar result is obtained under $E_x \to \infty$ ($\theta = \text{const}$) condition. It is related with a situation of $\theta \to 0$, since the potential barrier height $\Delta\Phi$ and the spacing between the system states $E_{ys} \approx \pm E_x$ increase with the driving field $E_x$.

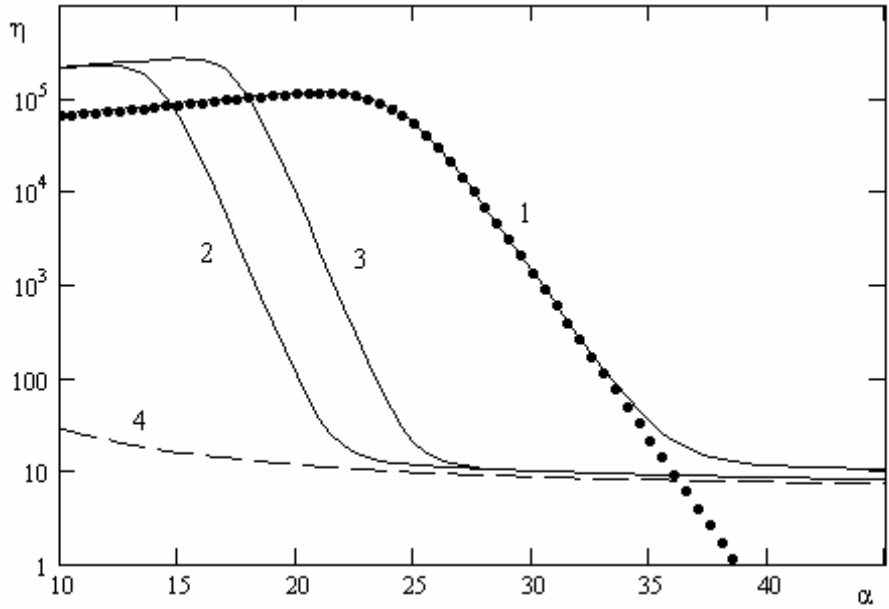

Fig. 1. Gain factor as a function of the parameter $\alpha$ ($\alpha = C_{11}/4\theta$):
$1 - E_x = 2$, $\Omega = 10^{-6}$; $2 - E_x = 2.5$, $\Omega = 10^{-6}$; $3 - E_x = 2.5$, $\Omega = 10^{-7}$;
$4 - E_x = 2.5$, $\Omega = 10^{-6}$.

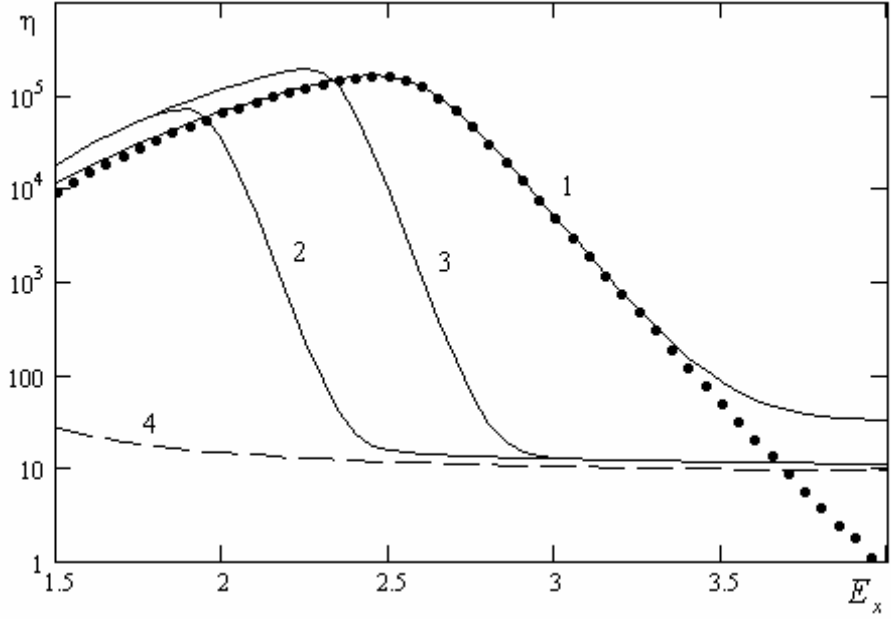

Fig. 2. Gain factor as a function of the driving field $E_x$: 1 – $\alpha = 10$, $\Omega = 10^{-5}$; 2 – $\alpha = 20$, $\Omega = 10^{-5}$; 3 – $\alpha = 20$, $\Omega = 10^{-7}$; 4 – $\alpha = 20$, $\Omega = 10^{-7}$.

## 4. Conclusion

In summary, a principal existence possibility is established of the stochastic resonance in 2SL. Owing to combined action of the noise and the driving electric field, the weak-signal gain factor can reach rather high values. The results obtained for $\eta = \eta(\Omega, T)$ agree well with general conclusions of the stochastic resonance theory. In our case, this phenomenon manifests two-parameter behavior, namely, both the noise power and the field $E_x$ affect the gain factor. At given temperature, the function $\eta = \eta(E_x)$ has a bell-like shape (see Fig. 2).

Note that NPT (and, correspondingly, stochastic resonance) in consideration is possible not only in 2SL, but also in "usual" bulk materials. As a sample of the latter, the conductors with bcc crystal lattice may be considered [1]. However, NPT in superlattices occurs at substantially lower driving electric field that facilitates, in principle, observation and employment of the phenomenon considered.

### Acknowledgements

The work was supported by the Russian Foundation of Fundamental Investigations (Project No. 02-02-16238).